\documentclass[aps,showpacs,preprintnumbers,amsmath, amssymb]{revtex4-1}

\usepackage{float}
\usepackage{graphics,epsfig}
\usepackage{graphicx}
\usepackage{dcolumn}
\usepackage{bm}

\begin{document}
\baselineskip=0.8 cm
\title{{\bf Holographic p-wave superfluid in the AdS soliton background with $RF^{2}$ corrections}}

\author{Yanmei Lv$^{1}$, Xiongying Qiao$^{1}$, Mengjie Wang$^{1}$\footnote{mjwang@hunnu.edu.cn}, Qiyuan Pan$^{1,2}$\footnote{panqiyuan@126.com}, Wei-Liang Qian$^{2,3,4}$\footnote{wlqian@usp.br}, and Jiliang Jing$^{1,2}$\footnote{jljing@hunnu.edu.cn}}
\affiliation{$^{1}$Key Laboratory of Low Dimensional Quantum Structures and Quantum Control of Ministry of Education, Synergetic Innovation Center for Quantum Effects and Applications, and Department of Physics, Hunan Normal University, Changsha, Hunan
410081, China}
\affiliation{$^{2}$Center for Gravitation and Cosmology, College of Physical Science and Technology, Yangzhou University, Yangzhou 225009, China}
\affiliation{$^{3}$ Escola de Engenharia de Lorena, Universidade de S\~ao Paulo, 12602-810, Lorena, SP, Brazil}
\affiliation{$^{4}$ Faculdade de Engenharia de Guaratinguet\'a, Universidade Estadual Paulista, 12516-410, Guaratinguet\'a, SP, Brazil}

\vspace*{0.2cm}
\begin{abstract}
\baselineskip=0.6 cm
\begin{center}
{\bf Abstract}
\end{center}

We investigate the holographic p-wave superfluid in the background metric of the AdS soliton with $RF^{2}$ corrections.
Two models, namely, the Maxwell complex vector field model and Yang-Mills theory, are studied in the above context by employing the Sturm-Liouville approach as well as the shooting method.
When turning on the spatial components of the gauge field, one observes that, in the probe limit, the inclusion of $RF^{2}$ corrections hinders the superfluid phase transition.
On the other hand, however, in the absence of the superfluid velocity, it is found that the $RF^2$ corrections lead to distinct effects for the two models. Regardless of either the $RF^2$ correction or the spatial component of the gauge field, the phase transition of the system is observed to be always of the second order.
Moreover, a linear relationship between the charge density and chemical potential is largely established near the critical point in both holographic superfluid models.

\end{abstract}
\maketitle
\newpage
\vspace*{0.2cm}

\section{Introduction}

The advent of the Anti-de Sitter/conformal field (AdS/CFT) theory~\cite{Maldacena,Gubser1998,Witten} opens up a new avenue for understanding of the pairing mechanism in the high $T_c$ superconductors which can not be described straightforwardly by the conventional BCS theory~\cite{BCS}.
The theory gives an account of a $d$-dimensional quantum field theory in the strong coupling regime in terms of a weakly coupled gravity theory.
The latter, also known as the bulk theory, is at least one dimension higher than the dual quantum field theory, often referred to as the boundary theory.
It has been suggested that, in the light of the AdS/CFT correspondence, the spontaneous $U(1)$ symmetry breaking in the bulk spacetime metric can be used to model the phase transition from the normal to superconducting state in the boundary theory dual to the gravitational system~\cite{GubserPRD78}.
The relevant transition is shown to exhibit the main characteristics of the s-wave superconductor~\cite{HartnollPRL101,HartnollJHEP12}.
These gravitational dual models are called holographic superconductors~\cite{HartnollRev,HerzogRev,HorowitzRev,CaiRev}.
Along this line of thought, by introducing an $SU(2)$ Yang-Mills field into the bulk, Gubser and Pufu constructed a holographic p-wave superconductor.
In their realization, a massive gauge boson is generated by the spontaneous breaking of the non-abelian gauge symmetry.
The latter is associated with one of the $SU(2)$ generators, and the resulting condensation is understood to be dual to the vector order parameter~\cite{GubserPufu}.
To go a step further, Cai~\emph{et al.} devised a new holographic p-wave superconductor model by considering a charged vector field in the Einstein-Maxwell theory with a negative cosmological constant.
The model can be viewed as a generalization of the $SU(2)$ model with a general mass and gyromagnetic ratio~\cite{CaiPWave-1,CaiPWave-2}.
In Refs.~\cite{DWaveChen,DWaveBenini}, the authors studied the properties of a charged massive spin two field propagating in the bulk and implemented the holographic d-wave superconductivity.
Further progress features the AdS soliton in the background metric of the bulk, as Nishioka \emph{et al.} demonstrated that the soliton might be unstable.
In particular, the formation of the scalar hair is impeded, and subsequently, a second-order phase transition takes place when the chemical potential is more significant than the critical value of $\mu_{c}$.
The resulting model is utilized to describe the transition between the insulator and superconductor~\cite{Nishioka-Ryu-Takayanagi}.

Most of the aforementioned works are featured by the Einstein-Maxwell theory coupled to a charged field on the gravity side.
According to the AdS/CFT correspondence, in the AdS spacetime, the curvature correction to the metric~\cite{Gregory,Pan-WangGB2010,NieZeng} and the higher derivative terms related to the gauge field~\cite{JS2010,WuCKW,SunWPJ2019} are expected to modify the dynamics of the dual field theory.
Interestingly enough, Myers~\emph{et al.} introduced a specific form of higher-order correction regarding the gauge field, namely, the $RF^{2}$ correction.
The latter arises from the Kaluza-Klein reduction of the five-dimensional Gauss-Bonnet gravity.
In particular, it has been argued that the correction term in question is universal in the sense that it can be used to produce the second-order equations of motion for both the gauge field and metric for any background~\cite{RCMyers}.
While studying the holographic properties of charged black holes with $RF^{2}$ corrections, Cai and Pang observed its impact on the DC conductivity~\cite{CaiPang}.
Also, by investigating the holographic s-wave superconductor with $RF^{2}$ corrections in the background of the AdS black hole, the authors of Ref.~\cite{ZPJCPL} found that the higher correction term facilitates the condensation of the scalar operator.
To be specific, a significant deviation from the standard value of the ratio of the gap frequency to the critical temperature was observed.
More recently, Lu~\emph{et al.} also constructed a holographic p-wave superconductor with $RF^{2}$ corrections.
Their approach is characterized by a Maxwell complex vector field in the five-dimensional AdS black hole and soliton background spacetimes~\cite{LuNPB2018}.
For the black hole background, it was observed that the $RF^{2}$ correction promotes the conductor/superconductor phase transition and causes the ratio of the gap frequency to the critical temperature to significantly deviate from the standard value.
On the contrary, for the soliton background, it was shown that the correction does not affect the critical chemical potential~\cite{LuNPB2018}.
In Ref. \cite{LuPLB2018}, the authors further extended the study to the Lifshitz gravity and obtained similar features for the effect of the $RF^2$ correction with respect to the holographic properties of the systems.

In this work, we examine the influence of the $RF^2$ corrections on the p-wave superfluid model.
According to the AdS/CFT correspondence, the holographic superfluid is realized by turning on the spatial components of the gauge field.
Special attention will be paid to the role of supercurrent, since it is an essential quantitiy concerning the study of superconductity in condensed matter systems~\cite{BasuMukherjeeShieh,HerzogKovtunSon,Peng2012,KuangLiuWang,Arean2010,AreanJHEP2010,SonnerWithers,ZengSZ,Amado2013,Zeng2013,Amado2014,LaiPJW2016,AriasLandea,GBSuperfluid,HuangPQJW}.
The calculations will be carried out for both the Maxwell complex vector field model and the Yang-Mills theory in the five-dimensional AdS Schwarzschild spacetime regarding the following soliton solution
\begin{eqnarray}\label{SchSoliton}
ds^2=-r^2dt^2+\frac{dr^2}{f\left(r\right)}+f\left(r\right)d\varphi^2+r^2(dx^2+dy^2),
\end{eqnarray}
with $f(r)=r^2(1-r_{s}^{4}/r^{4})$.
This solution does not possess any horizon but a conical singularity, corresponding to the tip of the soliton, at $r_{s}$.
One can avoid the singularity by imposing a period $\beta=\pi/r_{s}$ for the coordinate $\varphi$.
The motivation of the present study is to understand the influences of the $1/N$ or $1/\lambda$ (where $\lambda$ is the 't Hooft coupling) corrections on the holographic p-wave superfluid models.
As discussed in the following sections, in the probe limit where the backreaction of matter fields on the spacetime metric is neglected, the $RF^2$ corrections lead to qualitatively different effects on the superfluid phase transition in the two models with vanishing superfluid velocity.
With the presence of the superfluid velocity, on the other hand, similar features regarding the condensate of the vector operator are observed.
This indicates that one might make use of the $RF^2$ corrections to distinguish between the holographic p-wave superfluid state in the Maxwell complex vector field model and that in the Yang-Mills theory.

The present paper is organized as follows.
In Sec. II, we construct the holographic p-wave superfluid model with the $RF^2$ corrections via a Maxwell complex vector field model.
In the probe limit, an analytical method, Sturm-Liouville approach, is employed to study the effect of the $RF^2$ corrections on the superfluid phase transition.
The analysis is then complemented by a numerical method, namely, the shooting method.
In Sec. III, we extend the investigation to the holographic p-wave superfluid model with the $RF^2$ corrections to the Yang-Mills theory.
Finally, the last section is devoted to the discussions and concluding remarks.

\section{p-Wave superfluid of the Maxwell complex vector field}

In this section, we study the holographic p-wave superfluid phase transition with $RF^{2}$ corrections in the five-dimensional AdS soliton spacetime by considering the Maxwell complex vector field model~\cite{LuPLB2018,LuNPB2018}
\begin{eqnarray}\label{PWaveAtion}
S=\frac{1}{16\pi G}\int
d^{5}x\sqrt{-g}\left[-\frac{1}{4}F_{\mu\nu}F^{\mu\nu}+\mathcal{L}_{RF^{2}}
-\frac{1}{2}(D_\mu\rho_\nu-D_\nu\rho_\mu)^{\dag}(D^{\mu}\rho^{\nu}-D^{\nu}\rho^{\mu})
-m^2\rho_{\mu}^{\dag}\rho^{\mu}+iq\gamma_{0}\rho_{\mu}\rho_{\nu}^{\dag}F^{\mu\nu} \right],\nonumber \\
\end{eqnarray}
where the $RF^{2}$ correction term reads
\begin{eqnarray}
\mathcal{L}_{RF^{2}}=&\alpha (R_{\mu\nu\rho\lambda}F^{\mu\nu}F^{\rho\lambda}-4R_{\mu\nu}F^{\mu\rho}F^{\nu}_{\rho}
+RF^{\mu\nu}F_{\mu\nu}).
\end{eqnarray}
Here $D_\mu=\nabla_\mu-iqA_\mu$ is the covariant derivative and $F_{\mu\nu}=\nabla_{\mu}A_{\nu}-\nabla_{\nu}A_{\mu}$ is the strength of $U(1)$ field $A_\mu$. The coupling parameter $\alpha$ satisfies $-1/20\leq \alpha \leq 1/4$~\cite{RCMyers}, $q$ and $m$ are the charge and mass of the vector field $\rho_\mu$, respectively. The last term, proportional to $\gamma_{0}$, measures the interaction between the vector field $\rho_\mu$ and the gauge field $A_\mu$.

In order to investigate the possibility of DC supercurrent, according to Ref. \cite{LaiPJW2016}, we make use of the following {\it ansatz} for the matter fields
\begin{eqnarray}\label{PWaveAnsatz}
\rho_\mu dx^{\mu}=\rho_{x}(r)dx,~~A_\mu dx^{\mu}=A_t(r)dt+A_{\varphi}(r)d\varphi.
\end{eqnarray}
In the soliton background (\ref{SchSoliton}), one chooses $\rho_{x}(r)$, $A_t(r)$ and $A_{\varphi}(r)$ to be real functions.
Subsequently, one obtains the following equations of motion
\begin{eqnarray}\label{PWaveRhoxr}
\rho_{x}^{\prime\prime}+\left(\frac{1}{r}+\frac{f^\prime}{f}\right)\rho_{x}^{\prime}
-\frac{1}{f}\left(m^2+\frac{q^2A^2_\varphi}{f}-\frac{q^2A_t^2}{r^2}\right)\rho_{x}=0,
\end{eqnarray}
\begin{eqnarray}\label{PWaveAtr}
\left[1+\frac{8\alpha f}{r}\left(\frac{1}{r}+\frac{f'}{f}\right)\right]A_{t}''
+\left[\left(\frac{1}{r}+\frac{f'}{f}\right)+\frac{8\alpha}{r}\left(-\frac{f}{r^{2}}+\frac{2f'}{r}
+\frac{f'^{2}}{f}+f''\right)\right]A_{t}'-\frac{2q^{2}\rho_{x}^{2}}{r^{2}f}A_{t}=0,
\end{eqnarray}
\begin{eqnarray}\label{PWaveAvarphir}
\left(1+\frac{24\alpha f}{r^{2}}\right)A_{\varphi}''+\left[\frac{3}{r}+\frac{24\alpha f}{r^{2}}\left(\frac{1}{r}+\frac{f'}{f}\right)\right]A_{\varphi}'
-\frac{2q^{2}\rho_{x}^{2}}{r^{2}f}A_{\varphi}=0,
\end{eqnarray}
where the prime denotes the derivative with respect to $r$.
It is straightforward to show that Eqs.~(\ref{PWaveRhoxr}) and (\ref{PWaveAtr}) fall back to the case considered in Ref.~\cite{LuNPB2018} when the spatial component $A_{\varphi}$ is turned off.

Eqs.~(\ref{PWaveRhoxr}), (\ref{PWaveAtr}) and (\ref{PWaveAvarphir}) can be solved by using the following procedure.
At the tip $r=r_{s}$, the vector field $\rho_\mu$ and gauge field $A_\mu$ are required to be regular, and $A_{\varphi}(r_{s})=0$.
Also, as $r\rightarrow\infty$, the asymptotical behaviors of the solutions are
\begin{eqnarray}\label{PWInfinityCondition}
\rho_{x}=\frac{\rho_{x-}}{r^{\Delta_{-}}}+\frac{\rho_{x+}}{r^{\Delta_{+}}},~~A_t=\mu-\frac{\rho}{r^2},
~~A_\varphi=S_\varphi-\frac{J_\varphi}{r^2},
\end{eqnarray}
where $\Delta_{\pm}=1\pm\sqrt{1+m^2}$ are the characteristic exponents with the masses beyond the Breitenlohner-Freedman (BF) bound $m_{BF}^2=-1$. According to the AdS/CFT correspondence, $\mu$ and $S_\varphi$ are the chemical potential and superfluid velocity, while $\rho$ and $J_\varphi$ are the charge density and current in the dual field theory, respectively. Furthermore, we can interpret $\rho_{x-}$ and $\rho_{x+}$ as the source and vacuum expectation value of the
vector operator $O_{x}$ in the dual field theory.
Accordingly, we will impose boundary condition $\rho_{x-}=0$ to guarantee the spontaneous breaking of $U(1)$ gauge symmetry in the system.
For simplicity, we will use $\Delta$ to denote $\Delta_{+}$ in the following discussions.

It is straightforward to show that Eqs.~(\ref{PWaveRhoxr}), (\ref{PWaveAtr}) and (\ref{PWaveAvarphir}) are invariant with respect to the following scaling
transformations:
\begin{eqnarray}\label{PWSSymmetry}
&&r\rightarrow\lambda r,~~(t, \varphi, x,
y)\rightarrow\frac{1}{\lambda}(t, \varphi, x,
y),~~q\rightarrow q,~~(\rho_{x},A_{t},A_{\varphi})\rightarrow\lambda(\rho_{x},A_{t},A_{\varphi}),\nonumber \\
&&(\mu,S_\varphi)\rightarrow\lambda(\mu,S_\varphi),~~
(\rho,J_\varphi)\rightarrow\lambda^{3}(\rho,J_\varphi),~~\rho_{x+}\rightarrow\lambda^{1+\Delta}\rho_{x+},
\end{eqnarray}
where $\lambda$ is a positive number.
Subsequently, in what follows, we will present our results in terms of dimensionless quantities, which are invariant regarding Eq.~(\ref{PWSSymmetry}).

\subsection{Analytical approach by the Sturm-Liouville method}

We first use the Sturm-Liouville method~\cite{Siopsis,SiopsisB} to explore the effect of the $RF^{2}$ correction on the condensation as well as other critical phenomena of the system in the immediate vicinity of the critical chemical potential $\mu_{c}$.
The obtained solution provides an analytical understanding of the p-wave superfluid phase transition in the AdS soliton background.
For mathematical convenience, we will change the variable from $r$ to $z=r_{s}/r$ with the range $0<z<1$ in the following calculations.

We note that the vector field $\rho_{x}$ vanishes as long as one approaches the critical point $\mu_{c}$ from below. In this case, Eq.~(\ref{PWaveAtr}) can be simplified to read
\begin{eqnarray}\label{PWaveAtzCritical}
\left[1+8\alpha z^{3}f\left(\frac{1}{z}-\frac{f'}{f}\right)\right]A_{t}''+\left[\left(\frac{1}{z}+\frac{f'}{f}\right)+8\alpha z\left(3f-2zf'
-\frac{z^{2}f'^{2}}{f}-z^{2}f''\right)\right]A_{t}'=0,
\end{eqnarray}
where the prime denotes the derivative with respect to $z$, and the function $f$ is $f(z)=(1-z^{4})/z^2$.
The general solution of Eq.~(\ref{PWaveAtzCritical}) is found to be
\begin{eqnarray}\label{PWaveAtzCriSolution}
A_{t}=\mu+c_{1}\left[\ln\left(\frac{1+z^{2}}{1-z^{2}}\right)
+\frac{4\sqrt{2\alpha}}{\sqrt{1+24\alpha}}{\rm
ArcTan}\left(\frac{2\sqrt{2\alpha}z^{2}}{\sqrt{1+24\alpha}}\right)\right],
\end{eqnarray}
where $\mu$ and $c_{1}$ are the two constants of integration. By considering the Neumann-like boundary condition for the gauge field $A_{t}$, we must have $c_{1}=0$ to ensure that $A_{t}$ is finite at the tip $z=1$.
This is because the term between the square brackets is divergent at $z=1$.
Therefore we arrive at the solution of Eq.~(\ref{PWaveAtzCritical}), namely, $A_{t}(z)=\mu$ for $\mu<\mu_{c}$.

Similarly, as $\mu\rightarrow\mu_{c}$ from below, one finds, from Eq.~(\ref{PWaveAvarphir}), that
\begin{eqnarray}\label{PWaveAphizCritical}
(1+24\alpha z^{2}f)A_{\varphi}''+\left[-\frac{1}{z}+24\alpha z^{2}f\left(\frac{1}{z}+\frac{f'}{f}\right)\right]A_{\varphi}'=0.
\end{eqnarray}
By considering the boundary condition $A_\varphi(1)=0$, one obtains
\begin{eqnarray}\label{PWaveAtzCriticalSolution}
A_{\varphi}&=&S_{\varphi}\phi(z)\nonumber \\
&=&S_{\varphi}(1-z^{2})\left[1+8\alpha(1+z^{2}+z^{4})+\frac{192}{5}\alpha^{2}(z^{2}-1)(2+4z^{2}+6z^{4}+3z^{6})\right],
\end{eqnarray}
where we have neglected the terms of order $O(\alpha^{n})$ for $n\geq 3$.

Also, it is not difficult to show that, as $\mu\rightarrow\mu_{c}$, the vector field equation (\ref{PWaveRhoxr}) in terms of $z$ assumes the form
\begin{eqnarray}\label{PWRhozCriMotion}
\rho_{x}^{\prime\prime}+\left(\frac{1}{z}+\frac{f^\prime}{f}\right)\rho_{x}^\prime
+\left[\frac{1}{z^{2}f}\left(\frac{q\mu}{r_{s}}\right)^{2}-\frac{\phi^{2}}{z^{4}f^{2}}\left(\frac{qS_\varphi}{r_{s}}\right)^{2}
-\frac{m^{2}}{z^{4}f}\right]\rho_{x}=0.
\end{eqnarray}
By taking into account the asymptotical behavior of $\rho_x$ from Eq.~(\ref{PWInfinityCondition}), we make an {\it ansatz} of the following form~\cite{Siopsis}
\begin{eqnarray}\label{PWSLFz}
\rho_x(z)\sim \frac{\langle O_{x}\rangle}{r_{s}^{\Delta}}z^{\Delta}F(z),
\end{eqnarray}
where $F(z)$ is to be determined with the boundary condition $F(0)=1$.
The resulting equation of motion for $F(z)$ is found to be
\begin{eqnarray}\label{SLFzmotion}
(TF^{\prime})^{\prime}+T\left[U+V\left(\frac{q\mu}{r_{s}}\right)^{2}-W\left(\frac{qS_\varphi}{r_{s}}\right)^{2}\right]F=0,
\end{eqnarray}
with
\begin{eqnarray}\label{PWaveTUVWFu}
T=z^{1+2\Delta}f,~~
U=\frac{\Delta}{z}\left(\frac{\Delta}{z}+\frac{f^\prime}{f}\right)-\frac{m^2}{z^{4}f},~~V=\frac{1}{z^{2}f},~~ W=\frac{\phi^{2}}{z^{4}f^{2}}.
\end{eqnarray}
According to the standard procedure for the Sturm-Liouville eigenvalue problem~\cite{Gelfand-Fomin}, the first eigenvalue $\Lambda=q\mu/r_{s}$ can be obtained by using minimization principle in terms of Rayleigh quotient
\begin{eqnarray}\label{PWSLEigenvalue}
\Lambda^{2}=\left(\frac{q\mu}{r_{s}}\right)^{2}=\frac{\int^{1}_{0}T\left(F'^{2}-UF^{2}\right)dz}{\int^{1}_{0}T(V-k^{2}W)F^{2}dz},
\end{eqnarray}
where we have defined the dimensionless parameter $k=S_{\varphi}/\mu$.
It is noted that we have used the boundary condition $[T(z)F(z)F'(z)]|_{0}^{1}=0$ in order to derive the expression (\ref{PWSLEigenvalue}).
As a matter of fact, from Eq.~(\ref{PWaveTUVWFu}), we find that $T(1)\equiv0$, which leads to $T(1)F(1)F'(1)=0$.
Besides, the condition $T(0)F(0)F'(0)=0$ can also be satisfied automatically since the leading order contribution from $T(z)$ as $z\rightarrow0$ is $2\Delta-1=1+2\sqrt{1+m^2}\geq1$ with the mass $m^{2}\geq m_{BF}^2$.
This means that, as discussed in Refs.~\cite{HFLi,WangSPJ}, we need not impose any restrictions on $F'(z)$.
In other words, the Dirichlet boundary condition for the trial function, $F(0)=1$, is sufficient for the present purpose, and therefore we write
\begin{eqnarray}\label{TrialFunction}
F(z)=1-az,
\end{eqnarray}
with $a$ being a constant.
We note that Eq.~(\ref{TrialFunction}) is more appropriate than imposing an additional Neumann boundary condition, such as $F'(0)=0$.

As an example, we calculate the case for a given mass of the vector field $m^{2}=5/4$ together with $k=0.00$ and $\alpha=0.00$.
By choosing the form of the trial function as in Eq.~(\ref{TrialFunction}), we have
\begin{eqnarray}\label{Example}
\Lambda^{2}=\left(\frac{q\mu}{r_{s}}\right)^{2}=\frac{8100-14175a+6748 a^2}{48(21-35a+15a^2)},
\end{eqnarray}
whose minimum is found to be $\Lambda_{min}^{2}=7.757$ with $a=0.492$.
Thus, one finds the critical chemical potential to be $\Lambda_{c}=\Lambda_{min}=2.785(13)$, which is closer to the numerical value $\Lambda_{c}=2.784(99)$ obtained in Ref.~\cite{ZPJ2015}, in comparison with the analytical result $\Lambda_{c}=2.787$ shown in Table 2 of Ref.~\cite{LaiPJW2016}, deduced from the trial function $F(z)=1-az^{2}$ .
Similarly, when turning on the $RF^{2}$ correction and spatial component $A_{\varphi}$, for example, if one considers $\alpha=0.05$ and $k=0.25$, it is found that $\Lambda_{min}^{2}=8.000$ and $a=0.474$.
The latter subsequently lead to a critical chemical potential $\Lambda_{c}=\Lambda_{min}=2.828$.
In general, a similar procedure can be applied to obtain the value of the critical chemical potential analytically.
In Tables~\ref{PWaveTable} and \ref{PWaveTableM0}, we present the calculated critical chemical potential $\Lambda_{c}=q\mu_{c}/r_{s}$ for given $\alpha$, $k$, as well as the mass of the vector field.

\begin{table}[ht]
\begin{center}
\caption{\label{PWaveTable}
The calculated critical chemical potential $\Lambda_{c}=q\mu_{c}/r_{s}$ for the vector operator $O_{x}$ in the holographic p-wave superfluid of the Maxwell complex vector field model.
The results are obtained analytically by the Sturm-Liouville method (left column) and numerically by the shooting method (right column) for different $RF^2$ corrections strength $\alpha$, $k=S_{\varphi}/\mu$ and for a given mass of the vector field $m^{2}=5/4$.}
\begin{tabular}{c c c c c c c}
\hline
$\alpha$ &~~~~-0.03 &~~~~-0.01 &~~~~0    &~~~~0.01  &~~~~0.05  \\
\hline
$k=0.00$ &~~~~2.785~~2.785  &~~~~2.785~~2.785  &~~~~2.785~~2.785 &~~~~2.785~~2.785 &~~~~2.785~~2.785 \\
$k=0.25$ &~~~~2.790~~2.795  &~~~~2.800~~2.802  &~~~~2.805~~2.805  &~~~~2.811~~2.807  &~~~~2.828~~2.814  \\
$k=0.50$ &~~~~2.805~~2.825  &~~~~2.844~~2.856  &~~~~2.867~~2.867 &~~~~2.891~~2.877 &~~~~2.969~~2.906  \\
\hline
\end{tabular}
\end{center}
\end{table}

\begin{table}[ht]
\begin{center}
\caption{\label{PWaveTableM0}
The calculated critical chemical potential $\Lambda_{c}=q\mu_{c}/r_{s}$ for the vector operator $O_{x}$ in the holographic p-wave superfluid of the Maxwell complex vector field model.
The results are obtained analytically by the Sturm-Liouville method (left column) and numerically by the shooting method (right column) for different $RF^2$ corrections strength $\alpha$, $k=S_{\varphi}/\mu$, and for the massless vector field $m^{2}=0$.}
\begin{tabular}{c c c c c c c}
\hline
$\alpha$ &~~~~-0.03 &~~~~-0.01 &~~~~0    &~~~~0.01  &~~~~0.05  \\
\hline
$k=0.00$ &~~~~2.265~~2.265  &~~~~2.265~~2.265  &~~~~2.265~~2.265 &~~~~2.265~~2.265 &~~~~2.265~~2.265 \\
$k=0.25$ &~~~~2.271~~2.276  &~~~~2.280~~2.282  &~~~~2.285~~2.285  &~~~~2.290~~2.287  &~~~~2.305~~2.292  \\
$k=0.50$ &~~~~2.287~~2.308  &~~~~2.324~~2.336  &~~~~2.345~~2.345 &~~~~2.367~~2.353 &~~~~2.436~~2.378  \\
\hline
\end{tabular}
\end{center}
\end{table}

From Tables~\ref{PWaveTable} and \ref{PWaveTableM0}, for the case of $k=0$ with given $m$, the mass of the vector field, one finds that the critical chemical potential $\mu_{c}$ is independent of the strength of the $RF^2$ correction, $\alpha$.
It implies that the $RF^2$ correction does not affect the stability of the AdS soliton system, just as shown previously in Ref.~\cite{LuNPB2018}.
However, the situation is entirely different as we switch on the spatial component $A_{\varphi}$ of the gauge field.
For given $k$, for instance $k=0.25$ or $0.50$, we observe that the critical chemical potential $\mu_{c}$ increases as we increase the $RF^2$ correction in terms of $\alpha$.
This shows that, in general, a more significant $RF^2$ correction will make it harder for the holographic p-wave superfluid phase transition to be triggered.
Therefore, it is meaningful to further explore the impact of the $RF^2$ correction on the holographic p-wave superfluid, especially with nonvanishing spatial component $A_{\varphi}$.
For the given $\alpha$ and $m$, one finds that the critical chemical potential becomes more significant with increasing $k$.
This is in good agreement with the findings in Ref.~\cite{LaiPJW2016} and indicates that the spatial component of the gauge field hinders the superfluid phase transition.

Now, we move on to discuss the critical phenomena of the holographic p-wave system.
From Eq.~(\ref{PWaveAtr}), in the vicinity of the critical point one may expand $A_{t}(z)$ in terms of small $\langle O_{x}\rangle$ by
\begin{eqnarray}\label{PWAtExpand}
A_{t}(z)\sim\mu_{c}+\frac{2q^{2}\mu_{c}}{r_{s}^{2(1+\Delta)}}\langle
O_{x}\rangle^{2}\chi(z)+\cdots,
\end{eqnarray}
with the boundary condition $\chi(1)=0$ at the tip.
In turn, it provides the equation of motion for $\chi(z)$
\begin{eqnarray}\label{PWChiEoM}
(M\chi')'-z^{2\Delta-1}F(z)^{2}=0,
\end{eqnarray}
where we have defined
\begin{eqnarray}\label{PWMz}
M(z)=(1+24\alpha+8\alpha z^{4})zf.
\end{eqnarray}

By combining the asymptotic behavior of $A_{t}$ in Eq.~(\ref{PWInfinityCondition}) and Eq.~(\ref{PWAtExpand}), we may expand $A_{t}$ near $z\rightarrow0$ as
\begin{eqnarray}\label{PWPhiExp}
A_{t}(z)\simeq\mu-\frac{\rho}{r_{s}^{2}}z^2\simeq\mu_c
+2\mu_{c}\left(\frac{q\langle
O_{x}\rangle}{r_{s}^{1+\Delta}}\right)^{2}\left[\chi(0)+\chi^\prime(0)z+\frac{1}{2}\chi^{\prime\prime}(0)z^2+\cdot\cdot\cdot\right].
\end{eqnarray}
From the above equation one may derive the following relation by comparing the coefficients of the $z^{0}$ term on both sides
\begin{eqnarray}\label{PWOxExpre}
\frac{q\langle
O_{x}\rangle}{r_{s}^{1+\Delta}}=\frac{1}{\left[2\mu_c\chi(0)\right]^{\frac{1}{2}}}\left(\mu-\mu_c\right)^{\frac{1}{2}},
\end{eqnarray}
where $\chi(0)=c_{2}-\int^{1}_{0}M^{-1}\left[\int^{z}_{1}x^{2\Delta-1}F(x)^{2}dx\right]dz$ with the constant of integration $c_{2}$ being determined by the boundary condition of $\chi(z)$.
As an example, one obtains $\langle O_{x}\rangle\approx3.349(\mu-\mu_{c})^{1/2}$ and $a=0.474$ by assuming $k=0.25$, $\alpha=0.05$ and $m^{2}=5/4$, where, owing to the scaling symmetry shown in Eq. (\ref{PWSSymmetry}), we have also chosen $q=1$ and $r_{s}=1$ for simplicity.
From Eq.~(\ref{PWOxExpre}) one may conclude the scaling law $\langle O_{x}\rangle\sim\left(\mu-\mu_c\right)^{1/2}$.
This relation is valid in the immediate vicinity of the critical point and is independent of specific parameters of the $RF^2$ correction, the spatial component of the gauge field, and the mass of the vector field.
In other words, the phase transition of the holographic p-wave superfluid with $RF^{2}$ corrections in the Maxwell complex vector field model is of the second order, and the extracted critical exponent of the system is consistent with that of the mean-field value, $1/2$.

Furthermore, by examing the coefficients of the $z^1$ terms in Eq.~(\ref{PWPhiExp}), we observe that $\chi^\prime(0)\rightarrow 0$.
This behavior is actually consistent with the following relation from Eq.~(\ref{PWChiEoM})
\begin{eqnarray}\label{PWz1}
\left[\frac{\chi'(z)}{z}\right]\bigg|_{z\rightarrow0}=\chi''(0)
=-\frac{1}{(1+24\alpha)}\int^{1}_{0}z^{2\Delta-1}F(z)^{2}dz.
\end{eqnarray}

Moreover, by extracting the coefficients of the $z^2$ terms in Eq.~(\ref{PWPhiExp}), with the help of Eqs.~(\ref{PWOxExpre}) and (\ref{PWz1}), one finds
\begin{eqnarray}\label{PWRhoExpre}
\frac{\rho}{r_{s}^{2}}=-\left(\frac{q\langle
O_{x}\rangle}{r_{s}^{1+\Delta}}\right)^{2}\mu_{c}\chi^{\prime\prime}(0)=\Gamma(k,\alpha,m)(\mu-\mu_{c}),
\end{eqnarray}
with $\Gamma(k,\alpha,m)=[2(1+24\alpha)\chi(0)]^{-1}\int^{1}_{0}z^{2\Delta-1}F(z)^{2}dz$, which is a function of $k$, $\alpha$ and $m^{2}$.
For example, one obtains $\rho=1.069\left(\mu-\mu_c\right)$ by taking $a=0.474$, $k=0.25$, $\alpha=0.05$, and $m^{2}=5/4$, where, again, we have taken the freedom to scale the dimensional quantities and chosen $q=1$ and $r_{s}=1$ for simplicity.
We observe that the $RF^2$ correction, the spatial component of the gauge field, and the mass of the vector field will not alter Eq.~(\ref{PWRhoExpre}) except for the prefactor.
Therefore, we argue that, in the vicinity of the transition point, one may find a mostly linear relationship between the charge density and chemical potential, namely, $\rho\sim(\mu-\mu_{c})$ in the present model.

For the field $A_{\varphi}$, near $\mu_{c}$, Eq.~(\ref{PWaveAvarphir}) can be rewritten into
\begin{eqnarray}\label{PWEMAphizCritical}
(1+24\alpha z^{2}f)A_{\varphi}''+\left[-\frac{1}{z}+24\alpha z^{2}f\left(\frac{1}{z}+\frac{f'}{f}\right)\right]A_{\varphi}'
-\frac{2S_{\varphi}\phi(z)}{z^{2}f}\left(\frac{q\langle
O_{x}\rangle z^{\Delta}F}{r_{s}^{1+\Delta}}\right)^{2}=0,
\end{eqnarray}
which has a general solution
\begin{eqnarray}\label{PWEMAphizSolu}
A_\varphi=S_{\varphi}\phi(z)+S_{\varphi}\left(\frac{q\langle
O_{x}\rangle}{r_{s}^{1+\Delta}}\right)^{2}\int\frac{z}{1+24\alpha z^{2}f}\left[\int\frac{2x^{2\Delta-3}\phi(x)F(x)^{2}}{f(x)}
dx\right]dz.
\end{eqnarray}
By assuming $k=0.25$, $\alpha=0.05$, and $m^{2}=5/4$, as an example, we arrive at $A_\varphi=S_{\varphi}[\phi(z)+(0.0269-0.0288z^{2}+\cdot\cdot\cdot)\langle O_{x}\rangle^{2}$] with $a=0.474$, where, again, we have taken $q=1$ and $r_{s}=1$.
Obviously, the solution Eq.~(\ref{PWEMAphizSolu}) depends on the $RF^2$ correction.

\subsection{Numerical study by the shooting method}

In the previous section, we have made use of the Sturm-Liouville method to analytically investigate the properties of the holographic p-wave superfluid phase transition with $RF^{2}$ corrections in the vicinity of the transition point.
Now, we proceed to numerically study the holographic superfluid model by using the shooting method~\cite{HartnollRev,HerzogRev,HorowitzRev,CaiRev}. As the method is not restricted to the immediate vicinity of the critical chemical potential, the results obtained in the present section help to further explore the properties of the $RF^{2}$ correction on the condensation and critical phenomena of the system from a different perspective.
Moreover, it provides a means to compare the numerical results against the analytical ones, as well as to evaluate the accuracy and effectiveness of the expansion carried out concerning the Sturm-Liouville method.
Again, for convenience, we will make use of the scaling properties, Eq.~(\ref{PWSSymmetry}), to assume $q=1$ and $r_{s}=1$ when performing the numerical calculations.

\begin{figure}[ht]
\includegraphics[scale=0.626]{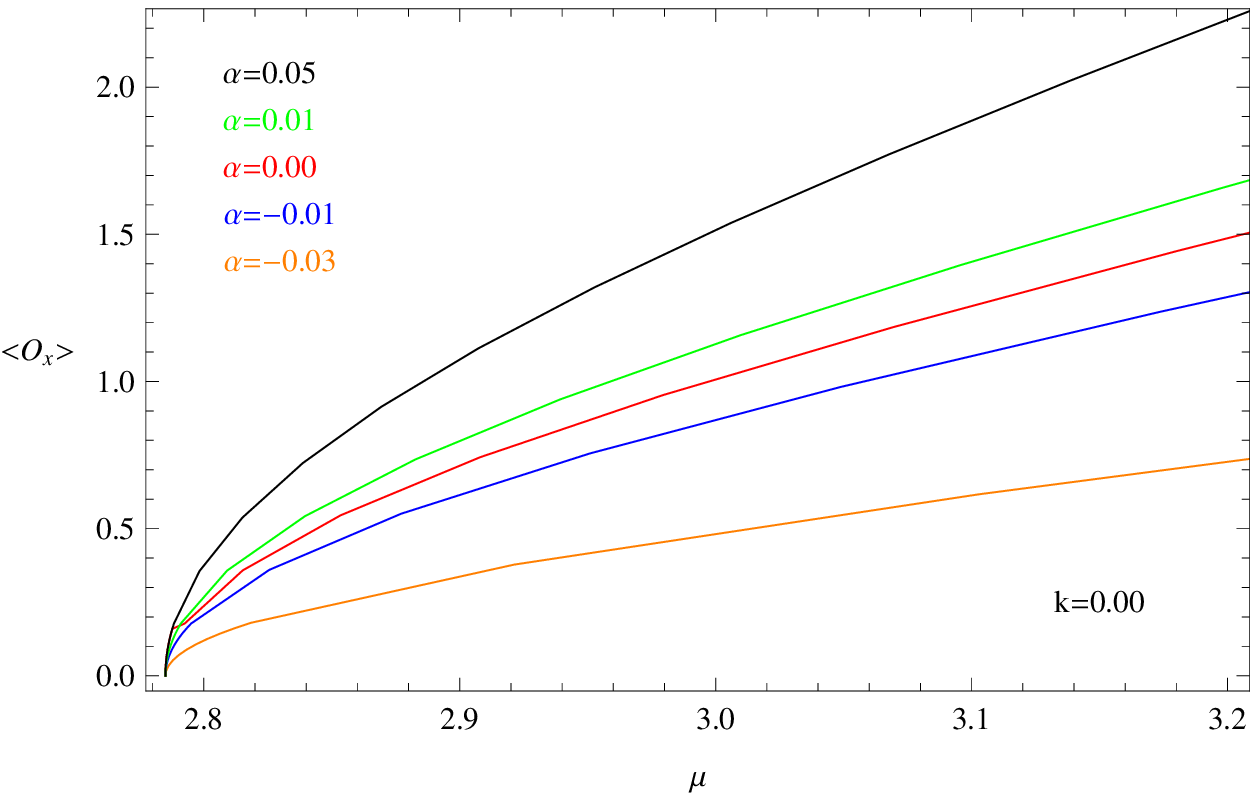}\;\;\includegraphics[scale=0.6]{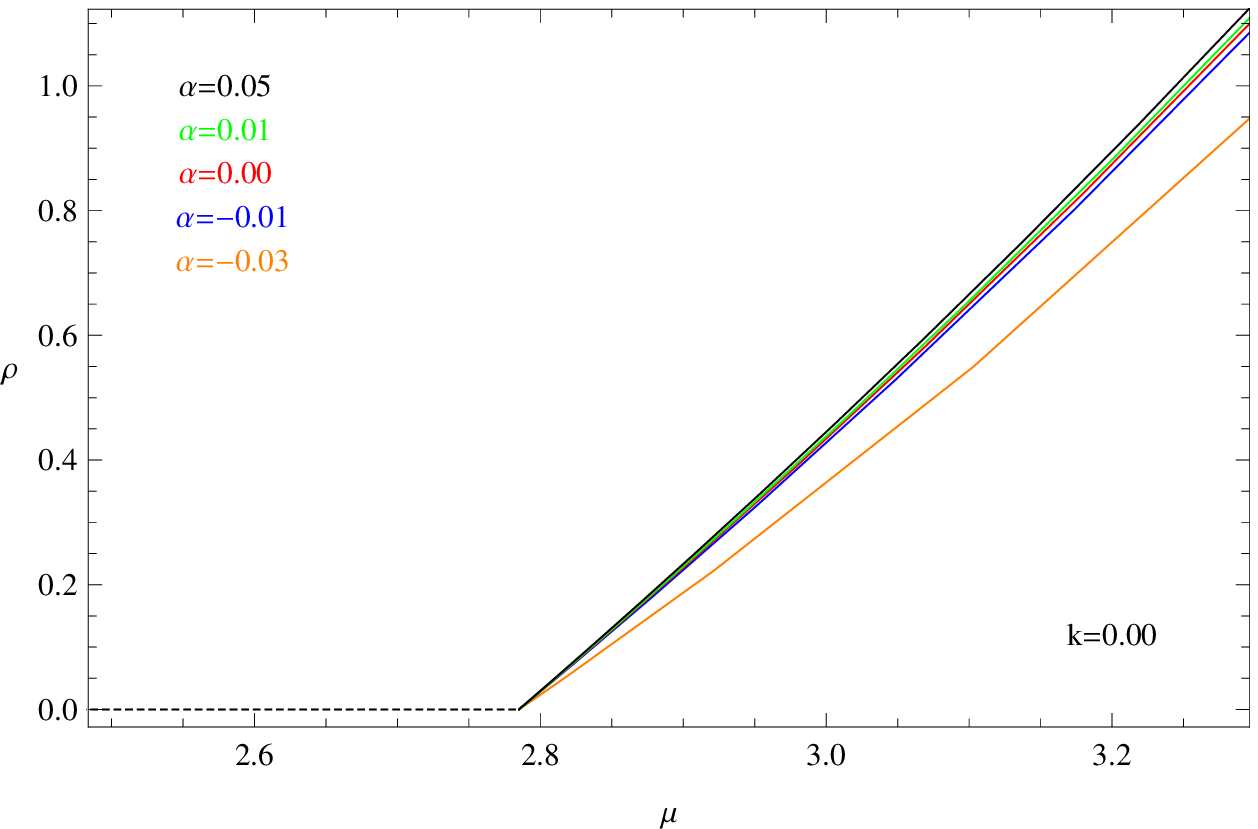}
\includegraphics[scale=0.626]{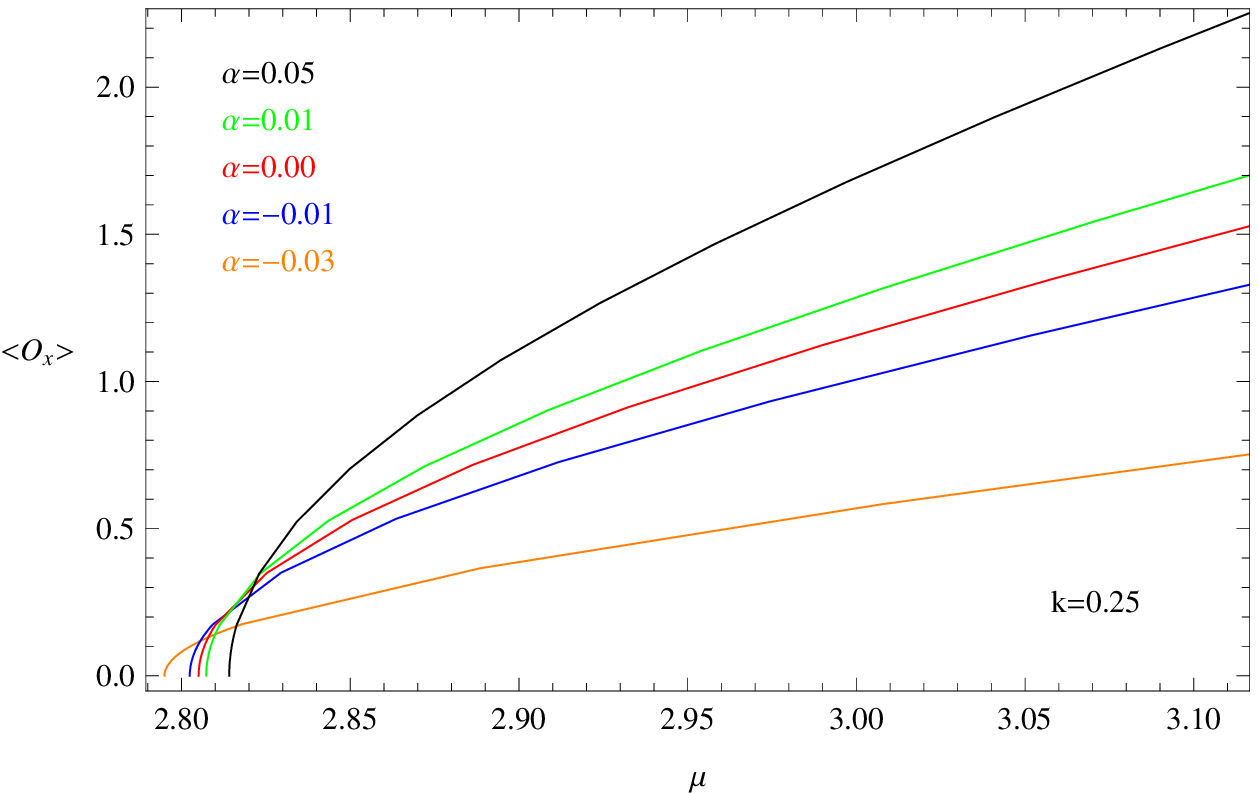}\;\;\includegraphics[scale=0.6]{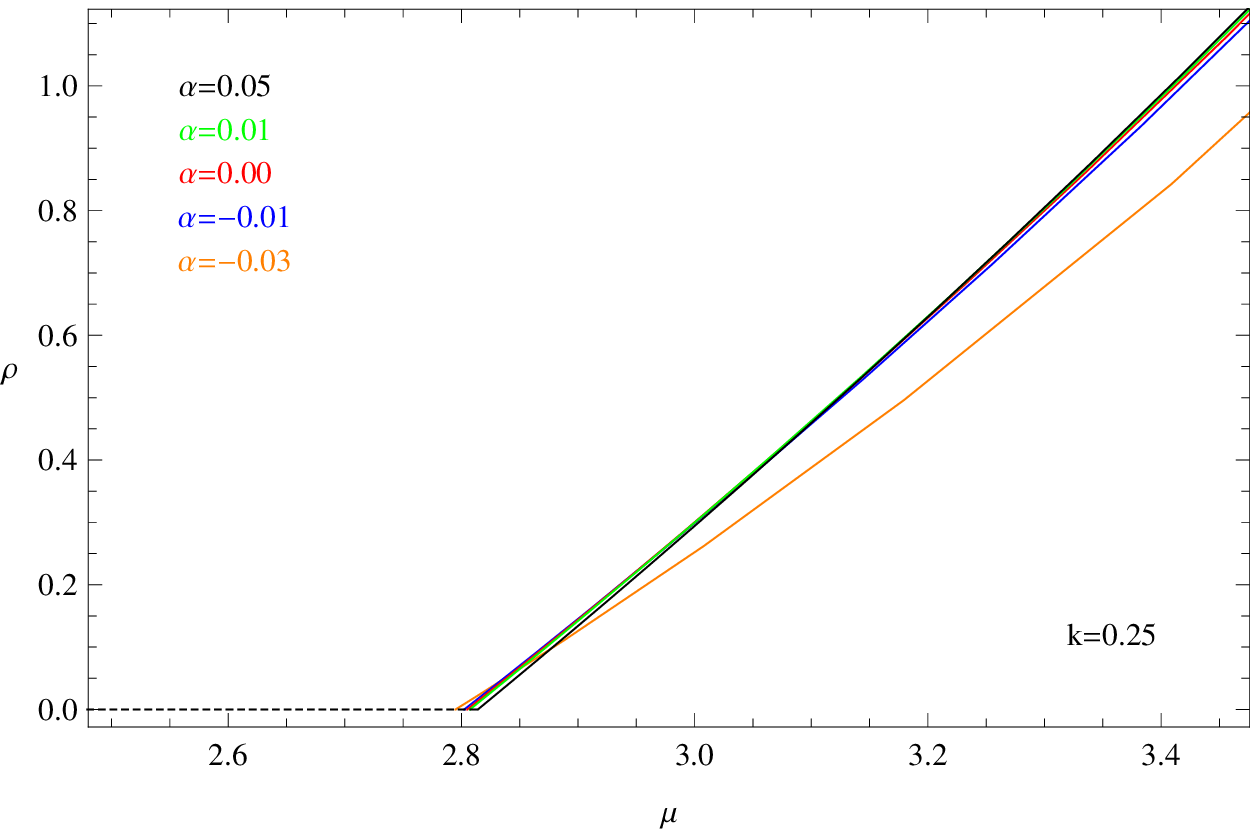}
\includegraphics[scale=0.626]{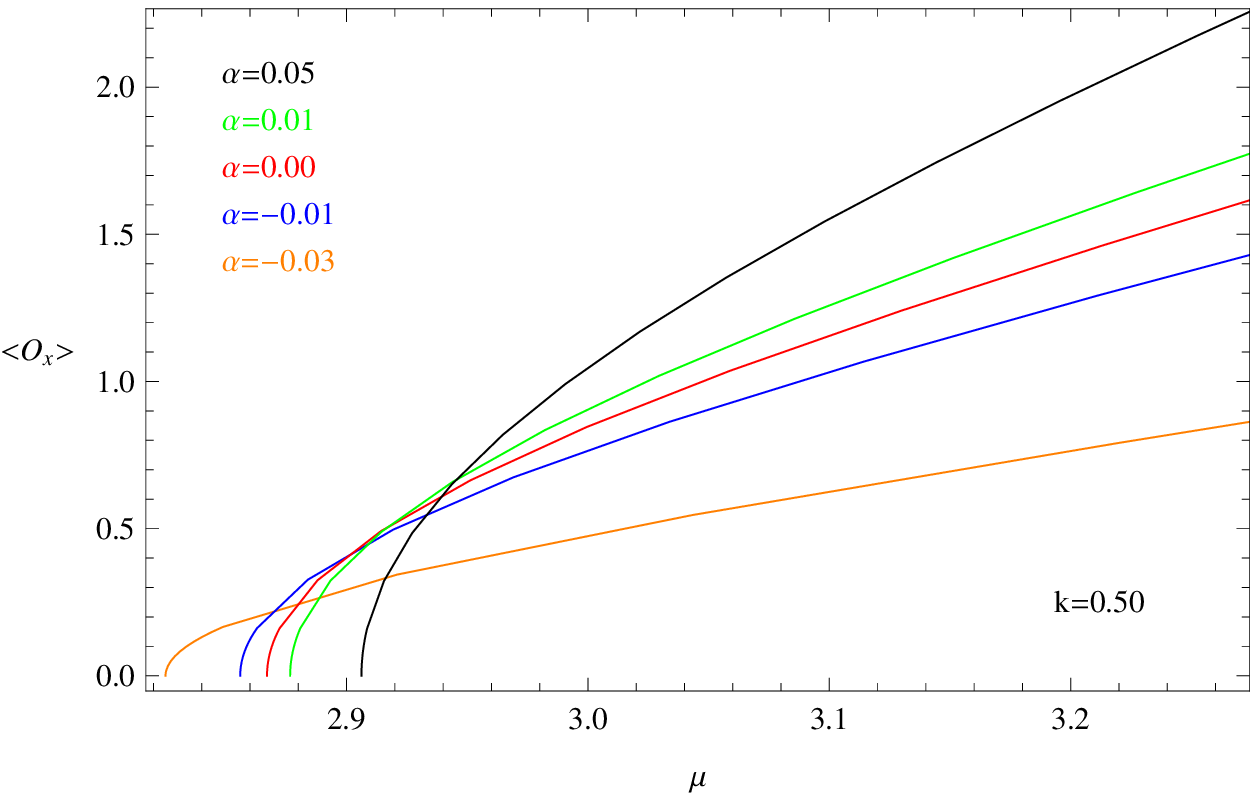}\;\;\includegraphics[scale=0.6]{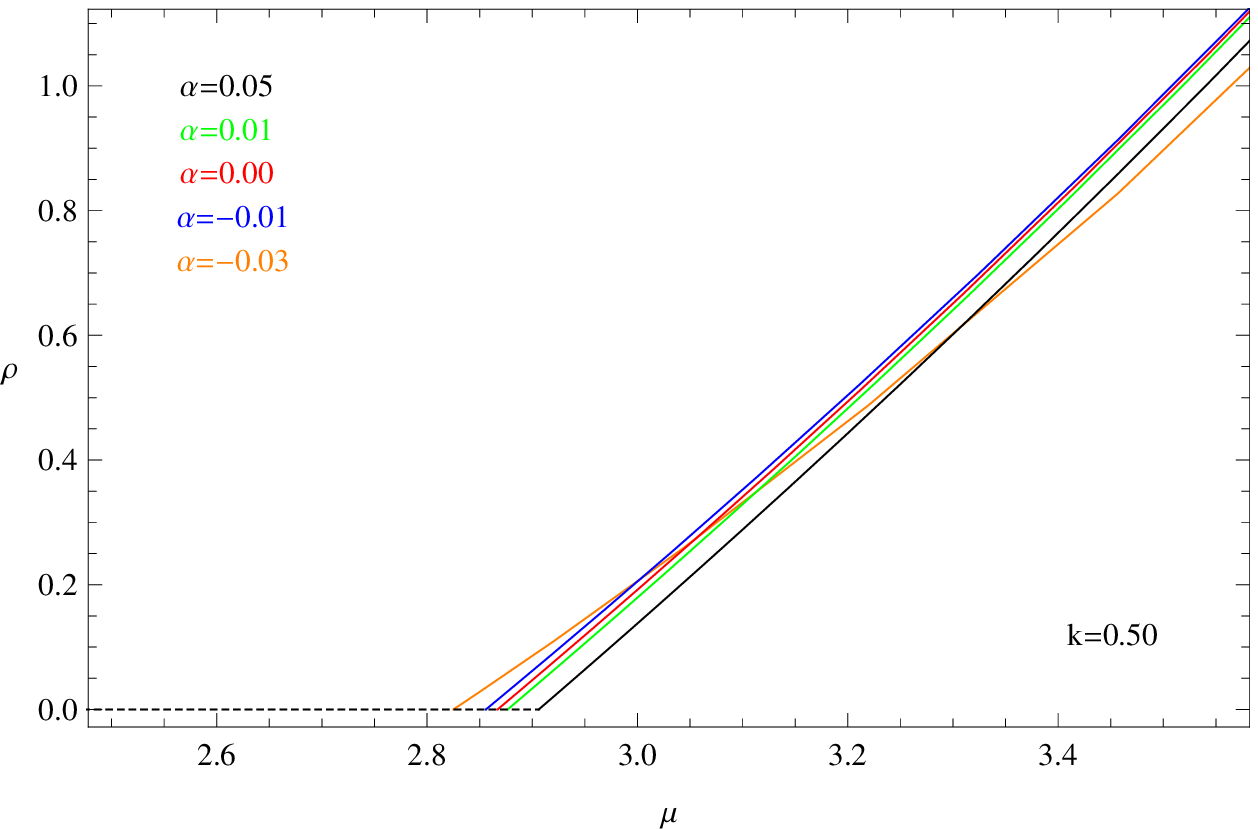}
\caption{\label{PWaveMCV}
(color online) The condensate $\langle O_{x}\rangle$ (left column) and charge density $\rho$ (right column) as functions of chemical potential $\mu$ for different values of $\alpha$ and $k=S_{\varphi}/\mu$ with $m^{2}=5/4$ in the holographic p-wave superfluid phase transition in the Maxwell complex vector field model.
In each plot, different curves correspond to $\alpha=-0.03$ (orange), $-0.01$ (blue), $0.00$ (red), $0.01$ (green) and $0.05$ (black) respectively.}
\end{figure}

By carrying out numerical integration from the tip to the infinity, one can solve the equations of motion (\ref{PWaveRhoxr}), (\ref{PWaveAtr}) and (\ref{PWaveAvarphir}).
On the left column of Fig.~\ref{PWaveMCV}, we plot the condensate of the vector operator $O_{x}$ as a function of the chemical potential for different values of $\alpha$, $k$, with given vector mass $m^{2}=5/4$.
It is shown that, the condensation occurs for $O_{x}$ with different values of $\alpha$ and $k$ if $\mu>\mu_{c}$.
As a comparison, we also present the critical chemical potential $\mu_{c}$ obtained numerically by using the shooting method in Table \ref{PWaveTable}.
It is noted that a satisfactory degree of agreement is achieved between the two methods.
This indicates that the Sturm-Liouville method is indeed powerful to analytically study the holographic superfluid models even with the presence of the $RF^{2}$ corrections.
It is confirmed that the critical chemical potential $\mu_{c}$ increases as $\alpha$ increases for the case where $k\neq0$, but it is mainly independent of $\alpha$ for the case where $k=0$, as can be observed both from Fig.~\ref{PWaveMCV} and Tables \ref{PWaveTable} and \ref{PWaveTableM0}.

On the other hand, from Fig.~\ref{PWaveMCV}, we find that, for all cases considered here, the vector operator $O_{x}$ is single-valued near the critical chemical potential and the condensate drops to zero continuously as the transition takes place.
By fitting these curves, we find that for small condensate, there is a square root behavior $\langle O_{x}\rangle\sim\left(\mu-\mu_c\right)^{1/2}$, which is also in good agreement with the analytical results discussed previously in Eq.~(\ref{PWOxExpre}).
As discussed before, this indicates the emergence of a second-order phase transition with the mean-field critical exponent $1/2$.
The $RF^2$ correction and the spatial component of the gauge field do not affect the result.

Furthermore, we present, in the right column of Fig.~\ref{PWaveMCV}, the charge density $\rho$ as a function of the chemical potential for different values of $\alpha$ and $k$ with given $m^{2}=5/4$.
For given $\alpha$ and $k$, we observe that the system is mostly described by the AdS soliton solution when $\mu$ is small, which can be interpreted as the insulator phase~\cite{Nishioka-Ryu-Takayanagi}.
When $\mu$ increases and reaches $\mu_{c}$, there is a phase transition, and the system transforms into the superfluid phase.
It is clearly shown that a linear relationship exists between the charge density and chemical potential near $\mu_{c}$, consistent with the analytical results discussed concerning Eq.~(\ref{PWRhoExpre}).
Here, we have numerically confirmed that the $RF^2$ correction and the spatial component of the gauge field do not affect the linear relation.

\section{p-Wave superfluid in the Yang-Mills theory}

In the previous section, we investigated the holographic p-wave superfluid with $RF^{2}$ corrections in the Maxwell complex vector field model.
Now, we extend our study of the holographic superfluid model to the non-abelian gauge field, namely, $SU(2)$ Yang-Mills theory with $RF^{2}$ corrections.
The action of the model reads
\begin{eqnarray}\label{YMaction}
S=\int
d^{5}x\sqrt{-g}\left[-\frac{1}{4\hat{g}^{2}}(F^{a}_{\mu\nu}F^{a\mu\nu}-4\mathcal{L}^{a}_{RF^{2}})\right],
\end{eqnarray}
with the $RF^2$ correction term
\begin{eqnarray}
\mathcal{L}^{a}_{RF^{2}}=\alpha (R_{\mu\nu\rho\lambda}F^{a\mu\nu}F^{a\rho\lambda}-4R_{\mu\nu}F^{a\mu\rho}F^{a\nu}_{\rho}
+RF^{a\mu\nu}F^{a}_{\mu\nu}),
\end{eqnarray}
where $\hat{g}$ is the Yang-Mills coupling constant and $F^{a}_{\mu\nu}=\partial_{\mu}A^{a}_{\nu}- \partial_{\nu}A^{a}_{\mu}+\varepsilon^{abc}A^{b}_{\mu}A^{c}_{\nu}$ is the strength of $SU(2)$ Yang-Mills field with the totally antisymmetric tensor $\varepsilon^{abc}$.
$A^{a}_{\mu}$ are the components of the mixed-valued gauge fields $A=A^{a}_{\mu}\tau^{a}dx^{\mu}$, where $\tau^{a}$ represent the three generators of the $SU(2)$ algebra which satisfy the commutation relation $[\tau^{a},\tau^{b}]=\varepsilon^{abc}\tau^{c}$.

Since we need a nonvanishing vector potential, we will adopt the following {\it ansatz} for the gauge fields~\cite{ZengSZ}
\begin{eqnarray}\label{YMansatz}
A(r)=A_{t}(r)\tau^{3}dt+\psi(r)\tau^{1}dx+A_{\varphi}(r)\tau^{3}d\varphi,
\end{eqnarray}
where the $U(1)$ subgroup of $SU(2)$ generated by $\tau^{3}$ is identified to be the electromagnetic gauge group.
Following Refs.~\cite{ZengSZ,GubserPRL2008}, we adopt the scenario of spontaneous symmetry breaking that the local $U(1)$ symmetry is broken down in the bulk, which corresponds to the holographic superfluid phase transition on the boundary. The latter is characterized by condensation in terms of the nonzero component $\psi(r)$ along the $x$-direction.
Subsequently, the vacuum state in question is no longer invariant with respect to the $U(1)$ symmetry, and therefore, according to the Higgs mechanism, a massive Higgs boson associated with $A_t$ is produced.
By making use of Eq.~(\ref{YMansatz}), one obtains the following equations of motion
\begin{eqnarray}\label{YMPWavePsir}
&&\left[1+\frac{8\alpha f}{r}\left(\frac{1}{r}+\frac{f'}{f}\right)\right]\psi''+\left[\left(\frac{1}{r}+\frac{f'}{f}\right)
+\frac{8\alpha}{r}\left(-\frac{f}{r^2}+\frac{2 f'}{r}+\frac{f'^2}{f}+f''\right)\right]\psi'\nonumber \\
&&+\left\{\left[1+4\alpha\left(\frac{2f'}{r}+f''\right)\right]\frac{A_{t}^{2}}{r^{2}f}
-\left[1+\frac{8\alpha f}{r}\left(\frac{1}{r}+\frac{f'}{f}\right)\right]\frac{A_{\varphi}^{2}}{f^{2}}\right\}\psi=0,
\end{eqnarray}
\begin{eqnarray}\label{YMPWaveAtr}
&&\left[1+\frac{8\alpha f}{r}\left(\frac{1}{r}+\frac{f'}{f}\right)\right]A_{t}''+\left[\left(\frac{1}{r}+\frac{f'}{f}\right)
+\frac{8\alpha}{r}\left(-\frac{f}{r^2}+\frac{2 f'}{r}+\frac{f'^2}{f}+f''\right)\right]A_{t}'\nonumber \\
&&-\left[1+4\alpha\left(\frac{2 f'}{r}+f''\right)\right]\frac{\psi^{2}}{r^{2}f}A_{t}=0,
\end{eqnarray}
\begin{eqnarray}\label{YMPWaveAphir}
\left(1+\frac{24\alpha f}{r^2}\right)A_{\varphi}''+\left[\frac{3}{r}+\frac{24\alpha f}{r^2}\left(\frac{1}{r}+\frac{f'}{f}\right)\right]A_{\varphi}'
-\left[1+\frac{8\alpha f}{r}\left(\frac{1}{r}+\frac{f'}{f}\right)\right]\frac{\psi^{2}}{r^{2}f}A_{\varphi}=0,
\end{eqnarray}
where the prime denotes the derivative with respect to $r$.
Obviously, in the case when $\alpha=0$, the two sets of equations of motion are equivalent if we further have $m^{2}=0$.
This can be readily verified by redefining the field by $\rho_{x}(r)=\psi(r)/\sqrt{2}$ in Eqs.~(\ref{PWaveRhoxr}), (\ref{PWaveAtr}) and (\ref{PWaveAvarphir}).
This result is essentially consistent with the arguments given by the authors of Ref.~\cite{CaiLLWPWave}, where they concluded that the complex vector field model could be viewed as a generalization of the $SU(2)$ Yang-Mills model.
However, for the present model, where the $RF^{2}$ correction has been introduced, such a conclusion does not hold.
As will be discussed below, the situation is entirely different when we consider the $RF^{2}$ corrections where $\alpha\neq0$.

We can solve the equations of motion (\ref{YMPWavePsir}), (\ref{YMPWaveAtr}) and (\ref{YMPWaveAphir}) by imposing the appropriate boundary conditions for the matter fields, i.e., the regularity condition at the tip $r=r_{s}$ and boundary behavior at the asymptotic boundary $r\rightarrow \infty$
\begin{eqnarray}\label{YMInfinityCondition}
\psi=\psi_{0}+\frac{\psi_{2}}{r^{2}},~~A_t=\mu-\frac{\rho}{r^2},
~~A_\varphi=S_\varphi-\frac{J_\varphi}{r^2},
\end{eqnarray}
where $\psi_{0}$ and $\psi_{2}=\langle O\rangle$ can be identified as a source and the expectation value of the dual operator.
We will use the asymptotic boundary condition $\psi_{0}=0$ since we are interested in the case where the condensation of the dual operator is spontaneous.

From Eqs.~(\ref{YMPWavePsir}), (\ref{YMPWaveAtr}) and (\ref{YMPWaveAphir}), one also finds that these equations are invariant regarding the following scaling transformation
\begin{eqnarray}
&&r\rightarrow\lambda r\,,\hspace{0.5cm}(t,\varphi,x,y)\rightarrow\frac{1}{\lambda}(t,\varphi,x,y)\,,\hspace{0.5cm}(\psi,A_{t},A_{\varphi})\rightarrow\lambda(\psi,A_{t},A_{\varphi})\,,\hspace{0.5cm}\nonumber \\&&(\mu,S_\varphi)\rightarrow\lambda(\mu,S_\varphi)\,,\hspace{0.5cm}(\rho,J_\varphi)\rightarrow\lambda^{3}(\rho,J_\varphi)\,,\hspace{0.5cm}\psi_{2}\rightarrow\lambda^{3}\psi_{2}\,,
\label{SLsymmetry-1}
\end{eqnarray}
where $\lambda$ is positive.

\subsection{Analytical approach by the Sturm-Liouville method}

We will closely follow the strategy utilized for the analysis regarding the Sturm-Liouville method in the previous section for the Maxwell complex vector field model.
First, we introduce the coordinate $z=r_{s}/r$.
By taking into consideration that the field $\psi=0$ as one approaches the transition point from below the critical chemical potential $\mu_{c}$, one may again derive the reduced equations of motion for the matter fields.
It is not difficult to show that, one actually arrives identical equations as those obtained in Eqs.~(\ref{PWaveAtzCritical}) and (\ref{PWaveAphizCritical}) for $A_{t}$ and $A_{\varphi}$, respectively.
This means that, as $\mu\rightarrow\mu_{c}$ from below the critical point, one obtains the physical solutions $A_{t}(z)=\mu$ and $A_{\varphi}(z)=S_{\varphi}\phi(z)$, identical to those of the Maxwell complex vector field model.
Thus, as $\mu\rightarrow\mu_{c}$, in terms of $z$, Eq.~(\ref{YMPWavePsir}) becomes
\begin{eqnarray}\label{YMpsiCriMotion}
&&\left[1+8\alpha z^{3}f\left(\frac{1}{z}-\frac{f'}{f}\right)\right]\psi''+\left[\left(\frac{1}{z}+\frac{f'}{f}\right)+8\alpha z\left(3f-2zf'-\frac{z^{2}f'^{2}}{f}-z^{2}f''\right)\right]\psi'\nonumber \\&&+\left\{(1+4\alpha z^{4}f'')\frac{1}{z^{2}f}\left(\frac{\mu}{r_{s}}\right)^{2}-\left[1+8\alpha z^{3}f\left(\frac{1}{z}-\frac{f'}{f}\right)\right]\frac{\phi^{2}}{z^{4}f^{2}}\left(\frac{S_{\varphi}}{r_{s}}\right)^{2}\right\}\psi=0,
\end{eqnarray}
where the function $\phi(z)$ has been defined in Eq.~(\ref{PWaveAtzCriticalSolution}).
When comparing with Eq.~(\ref{PWRhozCriMotion}) in the case of $S_{\varphi}=0$ and $m^{2}=0$, we find that Eq.~(\ref{YMpsiCriMotion}) is explicitly dependent on the coupling $\alpha$ even when $S_{\varphi}=0$.
This leads to the dependence of the critical chemical potential $\mu_{c}$ on the $RF^2$ correction in the holographic p-wave insulator/superconductor model ($k=0$) for the Yang-Mills theory.

Regarding the asymptotic behavior near the boundary, Eq.~(\ref{YMInfinityCondition}), we assume that $\psi$ takes the form
\begin{eqnarray}\label{YMWaveFz}
\psi(z)\sim \frac{\langle O\rangle}{r^{2}_{s}} z^{2}F(z),
\end{eqnarray}
where the trial function $F(z)$ with the boundary conditions $F(0)=1$ obeys equations of motion
\begin{eqnarray}\label{YMFzmotion}
(GF^{\prime})^{\prime}+G\left[Q+P\left(\frac{\mu}{r_{s}}\right)^{2}-W\left(\frac{S_{\varphi}}{r_{s}}\right)^{2}\right]F=0,
\end{eqnarray}
with
\begin{eqnarray}\label{YMGHPW}
G=(1+24\alpha+8\alpha z^{4})z^{5}f,~~
Q=-\frac{8(1+16\alpha+16\alpha z^{4})}{(1+24\alpha+8\alpha z^{4})f},~~
P=\frac{(1+24\alpha-8\alpha z^{4})}{(1+24\alpha+8\alpha z^{4})(1-z^{4})},
\end{eqnarray}
where $W(z)$ has been introduced in Eq.~(\ref{PWaveTUVWFu}). Solving the Sturm-Liouville eigenvalue problem \cite{Gelfand-Fomin}, we find
\begin{eqnarray}\label{YMMUC}
\Lambda^{2}=\left(\frac{\mu}{r_{s}}\right)^{2}=\frac{\int^{1}_{0}G(F'^{2}-QF^{2})dz}{\int^{1}_{0}G(P-Wk^{2})F^{2}dz} ,
\end{eqnarray}
which can be used to estimate the minimum eigenvalue of $\Lambda=\mu/r_{s}$.
One easily observes that $[G(z)F(z)F'(z)]|^1_{0}=0$, because of the fact that $G(1)\equiv0$ and $G(0)\equiv0$.
Therefore, similar to the Maxwell complex vector field model, we assume the trial function to be $F(z)=1-az$ with a constant $a$.

From the expression (\ref{YMMUC}), we can obtain the minimum eigenvalue of $\Lambda^{2}$ and the corresponding value of $a$ for different values of $k$ and $\alpha$.
For example, in the case of $k=0 $ and $\alpha=0$
\begin{eqnarray}
\Lambda^{2}=\left(\frac{\mu}{r_{s}}\right)^{2}=\frac{5(224-384a+189a^{2})}{14(15-24a+10a^{2})},
\end{eqnarray}
whose minimum is $\Lambda^{2}_{min}=5.132$ with $a=0.432$.
In comparison with the analytical result $\Lambda_{c}=\mu_{c}/r_{s}=2.267$ from the trial function $F(z)=1-az^{2}$ shown in Table 1 of Ref.~\cite{ZhaoPJ2013}, we have $\Lambda_{c}=2.265(47)$, which is closer to the numerical result $\Lambda_{c}=2.265(23)$.
In Table \ref{YMTable}, we present the calculated critical chemical potential $\Lambda_{c}$ for given $k$ and $\alpha$.

\begin{table}[ht]
\begin{center}
\caption{\label{YMTable}
The obtained critical chemical potential $\Lambda_{c}=\mu_{c}/r_{s}$ for the vector operator $O$ obtained analytically by the Sturm-Liouville method (left column) and numerically by the shooting method (right column).
The calculations are carried out with different $\alpha$, $k=S_{\varphi}/\mu$ in the holographic p-wave superfluid of the Yang-Mills field model.}
\begin{tabular}{c c c c c c c}
\hline
$\alpha$ &~~~~-0.03 &~~~~-0.01 &~~~~0    &~~~~0.01  &~~~~0.05  \\
\hline
$k=0.00$ &~~~~1.746~~1.704  &~~~~2.199~~2.199  &~~~~2.265~~2.265 &~~~~2.307~~2.306 &~~~~2.383~~2.383 \\
$k=0.25$ &~~~~1.747~~1.707  &~~~~2.212~~2.214  &~~~~2.285~~2.285 &~~~~2.333~~2.329  &~~~~2.433~~2.416  \\
$k=0.50$ &~~~~1.751~~1.714  &~~~~2.250~~2.260  &~~~~2.345~~2.345 &~~~~2.418~~2.402 &~~~~2.599~~2.524 \\
\hline
\end{tabular}
\end{center}
\end{table}

From Table \ref{YMTable}, for given $k$, one observes that the critical chemical potential $\mu_{c}$ increases with increasing $\alpha$.
This result agrees reasonably well with the findings in the Maxwell complex vector field model for $k\neq0$.
It indicates that a larger $RF^{2}$ correction hinders the phase transition.
Besides, for a given $\alpha$, $\mu_{c}$ becomes larger as $k$ increases, which is, again, consistent with the results in the Maxwell complex vector field model.
This implies that a nonvanishing spatial component of the gauge field makes the vector condensate harder to form~\cite{LaiPJW2016}.

Interestingly enough, for the case of $k=0$, one sees that $\mu_{c}$ is dependent on $\alpha$.
This is in contrast to the effect of the $RF^2$ correction for the Maxwell complex vector field model with $m^{2}=0$.
There, $\mu_{c}$ is independent of $\alpha$, as shown in Table~\ref{PWaveTableM0}.
Thus, we conclude that, in the case of $k=0$, the $RF^2$ corrections have entirely different effects between the insulator/superconductor phase transition of the Yang-Mills theory and that of the Maxwell complex vector field model.
This means that we can use the $RF^2$ corrections to distinguish between these two types of holographic superfluid models.

In order to analyze the critical phenomena of the system, we again expand $A_{t}(z)$ when $\mu\rightarrow\mu_{c}$ regarding $\langle O\rangle$ as
\begin{eqnarray}\label{YMEigenvalue}
A_{t}(z)\sim\mu_{c}+\frac{\mu_{c}}{r_{s}^{6}}\langle O\rangle^2\chi(z)+\cdots,
\end{eqnarray}
which gives rise to the following equation of motion in terms of $\chi(z)$
\begin{eqnarray}\label{YMXiEoM}
(M\chi')'-(1+24\alpha-8\alpha z^4)z^{3}F(z)^{2}=0,
\end{eqnarray}
where we have introduced the boundary condition $\chi(1)=0$ at the tip, and the function $M(z)$ has been defined in Eq.~(\ref{PWMz}).

By considering the asymptotic behavior and the expanded form of $A_{t}$ near $z\rightarrow0$, one finds
\begin{eqnarray}\label{YMPhiExpand}
A_{t}(z)\simeq\mu-\frac{\rho}{r_{s}^{2}}z^2\simeq\mu_c
+\mu_{c}\left(\frac{\langle O\rangle}{r^{3}_{s}}\right)^{2}\left[\chi(0)+\chi^\prime(0)z+\frac{1}{2}\chi^{\prime\prime}(0)z^2+\cdot\cdot\cdot\right].
\end{eqnarray}
By equating the coefficients of the $z^0$ term on both sides of the above equation, one gets
\begin{eqnarray}\label{YMOOExp}
\frac{\langle O\rangle}{r^{3}_{s}}=\frac{1}{\left[\mu_c\chi(0)\right]^{\frac{1}{2}}}\left(\mu-\mu_c\right)^{\frac{1}{2}},
\end{eqnarray}
where $\chi(0)=c_{3}-\int^{1}_{0}M^{-1} \left[\int^{z}_{1}(1+24\alpha-8\alpha x^4)x^{3}F(x)^{2}dx\right]dz$ with the constant of integration $c_{3}$ determined by the boundary condition of $\chi(z)$.
For given $k=0.25$ and $\alpha=0.05$, as an example, we find $\langle O\rangle\approx3.503(\mu-\mu_{c})^{1/2}$ with $a=0.498$, where, for simplicity, we have scaled the system to choose $r_{s}=1$.
Since Eq.~(\ref{YMOOExp}) is valid in genral, we obtain $\langle O\rangle\sim \left(\mu-\mu_c\right)^{1/2}$ near the critical point.
This indicates that the phase transition of the holographic superfluid with $RF^2$ corrections based on the Yang-Mills theory is of the second order.
Moreover, the critical exponent of the system attains the mean-field value 1/2.
It is noted that the $RF^2$ correction and the spatial component of the gauge field do not influence the result.

In addition, by comparing the coefficients of the $z^2$ terms on both sides of Eq.~(\ref{YMPhiExpand}), we have
\begin{eqnarray}\label{YMExpRho}
\frac{\rho}{r_{s}^{2}}=-\frac{1}{2}\left(\frac{\langle
O\rangle}{r^{3}_{s}}\right)^2\mu_c\chi^{\prime\prime}(0)=\Gamma(k,\alpha)(\mu-\mu_{c}),
\end{eqnarray}
with $\Gamma(k,\alpha)=[2(1+24\alpha)\chi(0)]^{-1} \int^{1}_{0}(1+24\alpha-8\alpha z^4)z^{3}F(z)^{2}dz$.
This is a function of the parameters $k$ and $\alpha$.
For example, in the case of $k=0.25$ with $\alpha=0.05$, we obtain $\rho=1.270\left(\mu-\mu_c\right)$ with $a=0.498$, where we have again scaled the system to have $r_{s}=1$, for simplicity.
Obviously, in the vicinity of the critical point, the linear relationship between the charge density and chemical potential $\rho\sim(\mu-\mu_{c})$ is valid in general for the holographic superfluid model of the Yang-Mills theory.

Similarly, when $\mu\rightarrow\mu_{c}$, Eq.~(\ref{YMPWaveAphir}) for the field $A_{\varphi}$ can be rewritten into
\begin{eqnarray}\label{YMEMAphizCri}
(1+24\alpha z^{2}f)A_{\varphi}''+\left[-\frac{1}{z}+24\alpha z^{2}f\left(\frac{1}{z}+\frac{f'}{f}\right)\right]A_{\varphi}'
-\left[1+8\alpha z^{3}f\left(\frac{1}{z}-\frac{f'}{f}\right)\right]\frac{S_{\varphi}\phi(z)}{z^{2}f}\left(\frac{\langle
O\rangle z^{2}F}{r^{3}_{s}}\right)^{2}=0.
\end{eqnarray}
Hence we finally find
\begin{eqnarray}\label{YMEMAphizSolution}
A_\varphi=S_{\varphi}\phi(z)+S_{\varphi}\left(\frac{\langle
O\rangle}{r_{s}^{3}}\right)^{2}\int\frac{z}{1+24\alpha z^{2}f(z)}\int\left\{1+8\alpha x^{3}f(x)\left[\frac{1}{x}-\frac{f'(x)}{f(x)}\right]\right\}\frac {x\phi(x)F(x)^{2}}{f(x)}dxdz.
\end{eqnarray}
As an example, for given $k=0.25$ and $\alpha=0.05$, we have $A_\varphi=S_{\varphi}[\phi(z)+(0.0377-0.0570z^{2}+\cdot\cdot\cdot)\langle O\rangle^{2}$] with $a=0.498$ and $r_{s}=1$.
It is consistent with the previous findings for the Maxwell complex vector field model.

\subsection{Numerical study by the shooting method}

\begin{figure}[ht]
\includegraphics[scale=0.616]{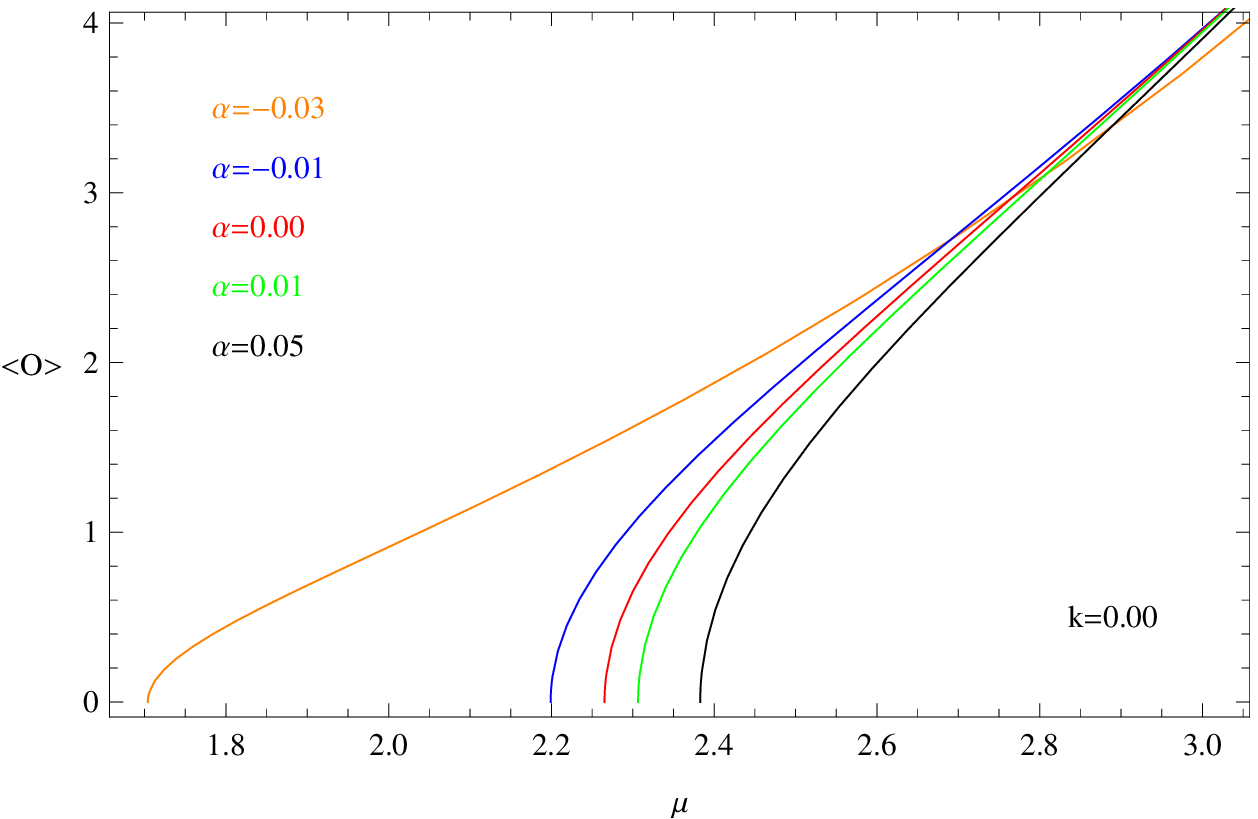}\;\;\includegraphics[scale=0.6]{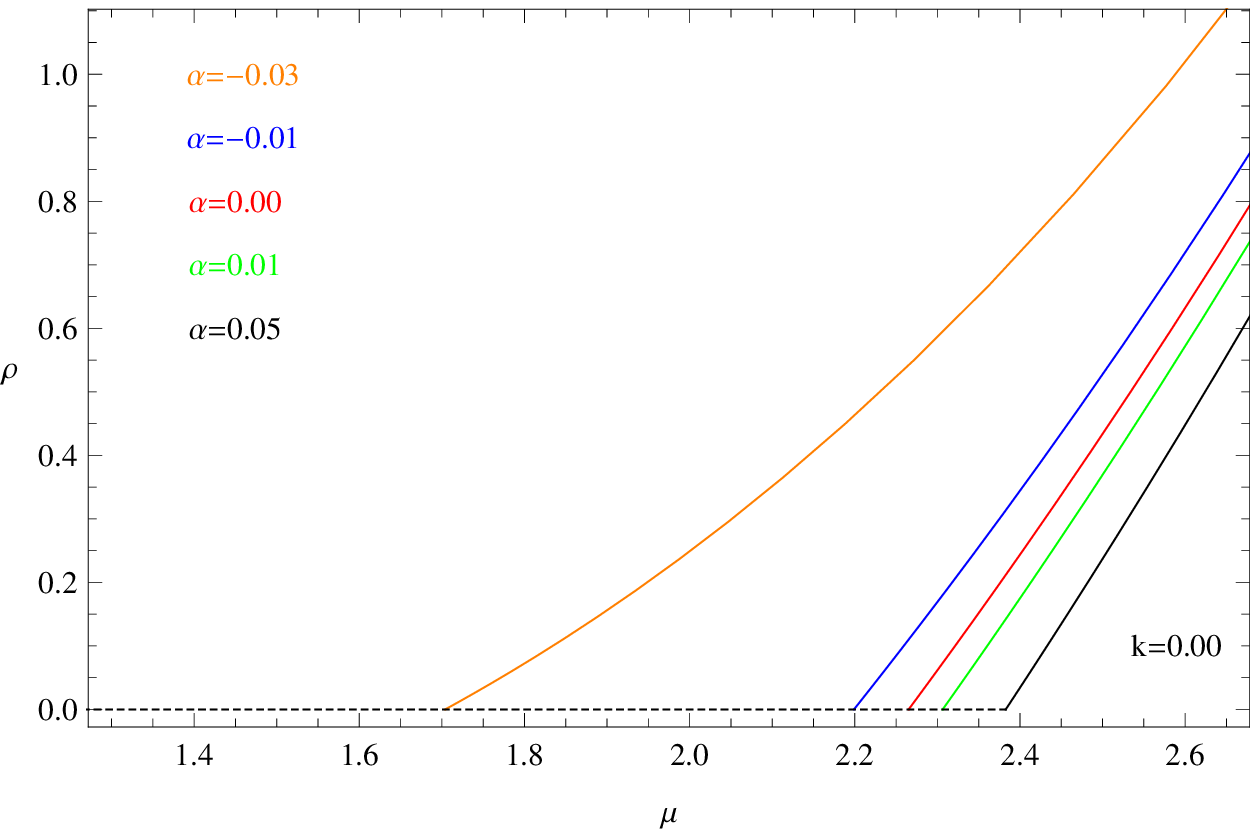}
\includegraphics[scale=0.616]{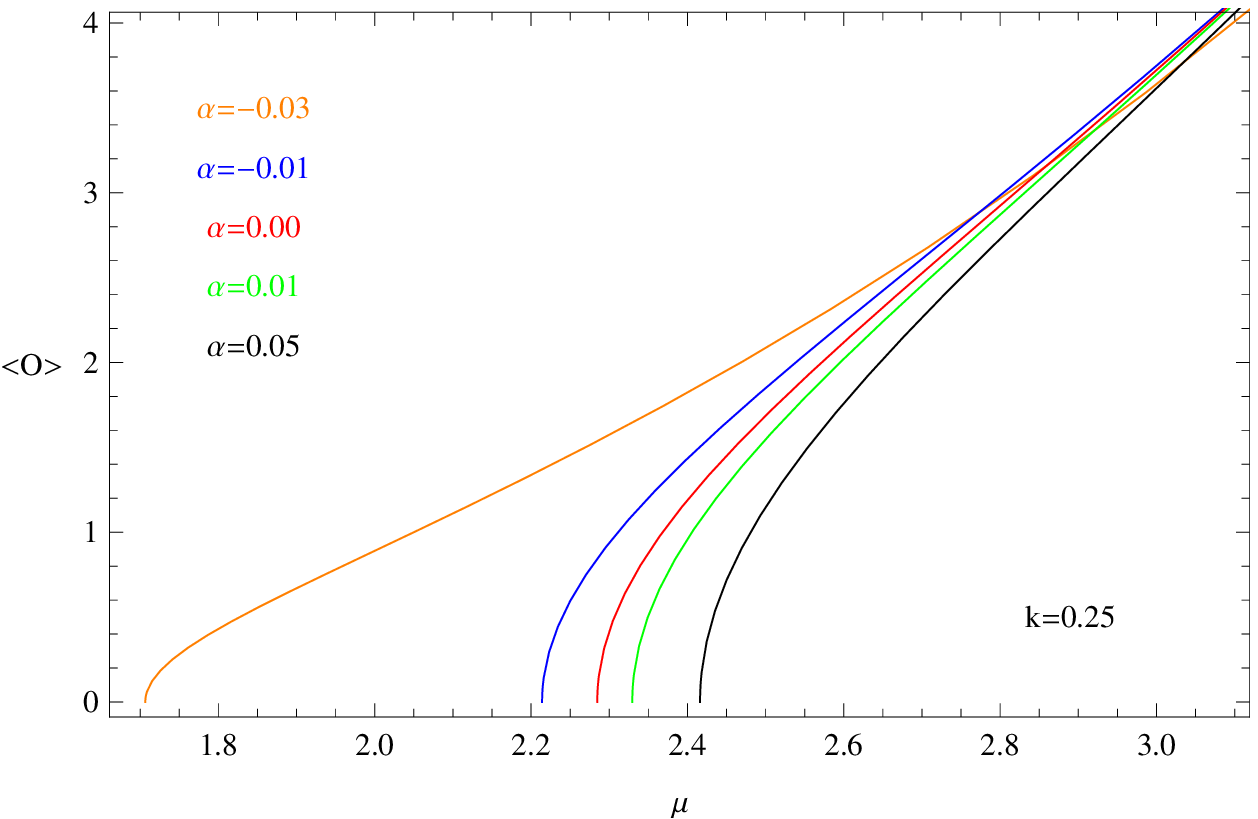}\;\;\includegraphics[scale=0.6]{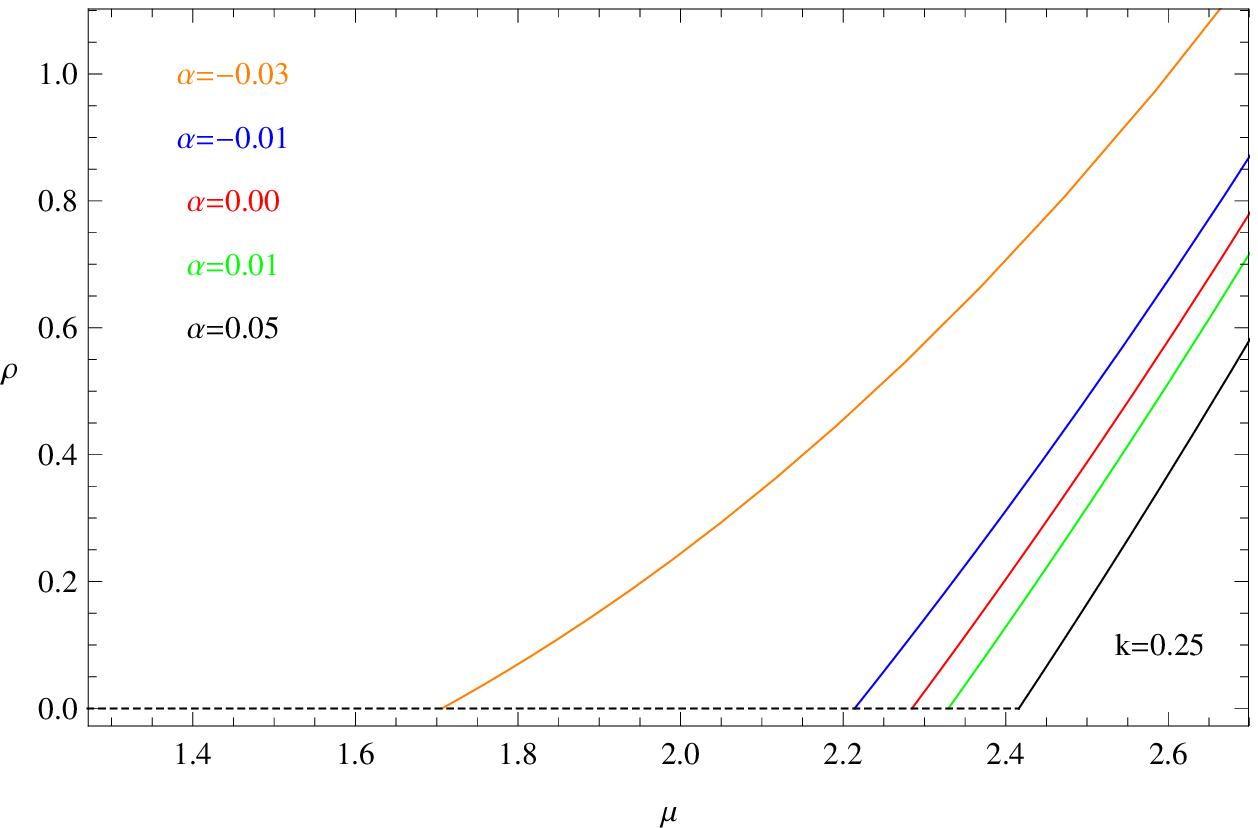}
\includegraphics[scale=0.616]{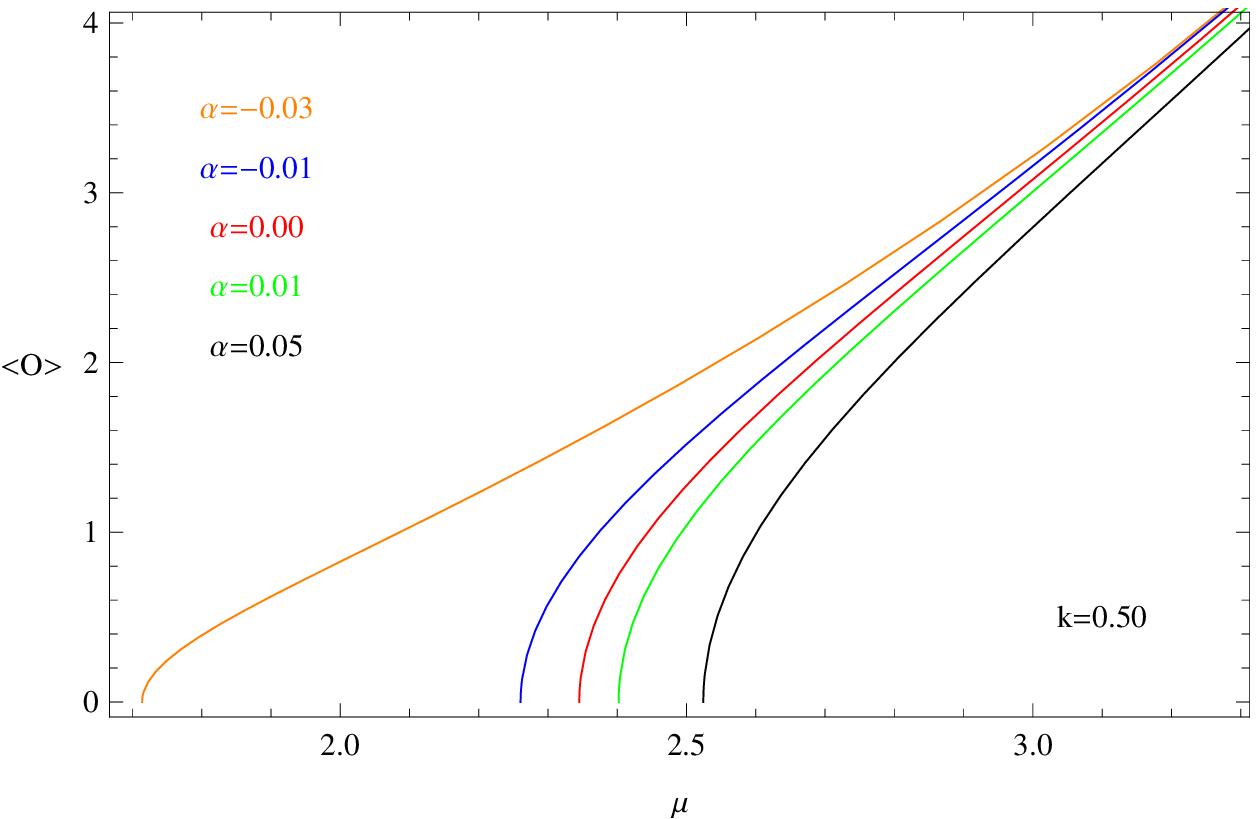}\;\;\includegraphics[scale=0.6]{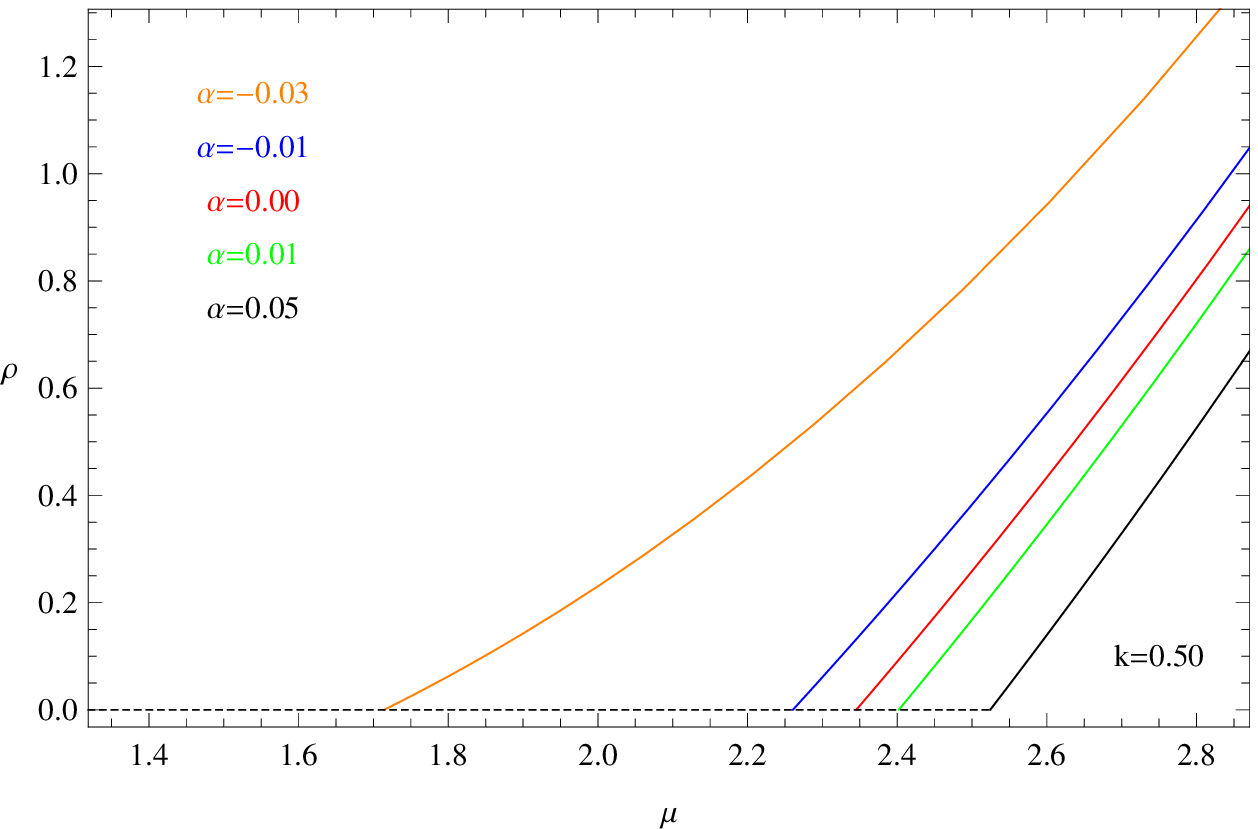}
\caption{\label{YMfigure}
(color online) The condensate $\langle O\rangle$ (left column) and charge density $\rho$ (right column) as functions of the chemical potential $\mu$ for different values of $\alpha$ and $k=S_{\varphi}/\mu$ in the holographic p-wave superfluid phase transition of the Yang-Mills theory.
In each plot, different curves correspond to $\alpha=-0.03$ (orange), $-0.01$ (blue), $0.00$ (red), $0.01$ (green) and $0.05$ (black) respectively.}
\end{figure}

In this section, the shooting method is employed to solve the equations of motion (\ref{YMPWavePsir}), (\ref{YMPWaveAtr}) and (\ref{YMPWaveAphir}).
In our numerical calculations, $r_{s}=1$ is chosen for convenience.
The results are presented in Fig.~\ref{YMfigure}.
In the left column, we show the condensate of the vector operator $O$ as a function of the chemical potential.
It is found that a phase transition occurs as $\mu$ increases and reaches $\mu_{c}$. Subsequently, the AdS soliton transforms into the superfluid phase.
The transition point is dependent on specific values of $\alpha$ and $k$.
Also, the conclusion that $\alpha$ affects the value of $\mu_{c}$ can also be drawn from the results presented in Table \ref{YMTable}.
Moreover, from Table \ref{YMTable}, it is observed that the numerical results (shown in the right column) agree well with the analytical ones derived from the Sturm-Liouville method (shown in the left column).
From Fig.~\ref{YMfigure} and Table \ref{YMTable}, we confirm that for given $k$, the critical chemical potential increases with increasing $\alpha$, previously obtained in the last section.
It implies that a larger $RF^{2}$ correction will make the vector condensate harder to take place.

From the left column of Fig.~\ref{YMfigure}, one also finds that the transition is of the second order and the condensate approaches zero according to the form $\langle O\rangle\sim(\mu-\mu_{c})^{\beta}$ with the critical exponent $\beta=1/2$ in accordance with the mean-field theory. For all cases considered here, this result is independent of either the $RF^2$ correction or the spatial component of the gauge field.
This is in good agreement with the analytical result discussed previously in Eq.~(\ref{YMOOExp}).

From the right column of Fig.~\ref{YMfigure}, we confirm numerically a linear relationship between the charge density and chemical potential in the vicinity of $\mu_{c}$, namely, $\rho\sim(\mu-\mu_{c})$.
For all the cases considered here, it agrees well with the analytical one derived in Eq.~(\ref{YMExpRho}).
The $RF^2$ correction and the spatial component of the gauge field do not affect the observed linearity.

\section{Conclusions}

In order to understand the influences of the $1/N$ or $1/\lambda$ corrections on the vector condensate in the holographic p-wave superfluid, we have investigated the role of the $RF^{2}$ corrections in the AdS soliton background for both the Maxwell complex vector field model and Yang-Mills theory.
In the probe limit, the calculations were carried out by employing the analytical Sturm-Liouville method as well as the numerical shooting method.
The results obtained by the two distinct methods were found to agree with each other to a satisfactory degree.
By turning on the spatial components of the gauge field, we observed that the critical chemical potential $\mu_{c}$ increases as the strength of the $RF^{2}$ correction, $\alpha$, increases.
This indicates that a larger $RF^{2}$ correction hinders the superfluid phase transition in both models. However, in the absence of the superfluid velocity, we noted that the transition point regarding $\mu_{c}$ is insensitive to $\alpha$ for the case of the Maxwell complex vector field model, while it is sensitively dependent on $\alpha$ in the Yang-Mills theory. In other words, the $RF^2$ corrections imply very different effects for the two different models.
This feature might be attributed to the intrinsic difference between the two models in question.
To be more specific, although both models effectively involve vector field, as well as electromagnetic field degrees of freedom and their condensate, the mass of the vector field is obtained by an explicit symmetric breaking in the complex vector model, while the relevant degree of freedom is derived through spontaneous symmetric breaking of $SU(2)$ gauge in the Yang-Mills theory. Moreover, by taking the mass of the vector field in the Maxwell complex vector field model, as well as the $RF^2$ correction, to be zero, one can readily show that the two sets of equations of motion for the two models are equivalent.
This result is similar to what has been pointed out in Ref.~\cite{CaiLLWPWave}. In this context, the authors of Ref.~\cite{CaiLLWPWave} argued that the complex vector model can be seen as a generalization of the Yang-Mills model of holographic superconductor/superfluid.
We understand that the above characteristics can be utilized to distinguish between these two types of superfluid models. Furthermore, for both models, we showed that the phase transition of the system is of the second order, and a linear relationship is found between the charge density and chemical potential in the vicinity of the critical point.
The presence of the $RF^2$ correction or the spatial component of the gauge field does not modify this result.
The present work is carried out in the framework of the probe limit, although such approximation is known to capture the essential features of the problem while significantly simplifies the mathematical formulation, it would still be of great interest to extend the study to take into consideration of the backreaction.
We plan to continue the work in a future study.

\begin{acknowledgments}

This work was supported by the National Natural Science Foundation of China under Grant Nos. 11775076, 11875025, 11705054 and 11690034; Hunan Provincial Natural Science Foundation of China under Grant No. 2016JJ1012;
as well as Brazilian funding agencies Funda\c{c}\~ao de Amparo \`a Pesquisa do Estado de S\~ao Paulo (FAPESP),
Conselho Nacional de Desenvolvimento Cient\'{\i}fico e Tecnol\'ogico (CNPq), and Coordena\c{c}\~ao de Aperfei\c{c}oamento de Pessoal de N\'ivel Superior (CAPES).

\end{acknowledgments}

\end{document}